\documentclass[runningheads]{llncs}
\RequirePackage{graphicx}
\RequirePackage{listings}
\RequirePackage{amsmath}
\RequirePackage{xcolor}
\RequirePackage[hidelinks]{hyperref}
\RequirePackage{cleveref}
\lstdefinelanguage{isabelle}{
  morekeywords={theorem,theorems,corollary,lemma,lemmas,locale,begin,end,fixes,assumes,shows,
    constrains , definition, where, apply, done,unfolding, primrec, using, by, for, uses,
    schematic_lemma, concrete_definition, prepare_code_thms, export_code, datatype,
    proof, next, qed, show, have, hence, thus, interpretation, fix, context, fun, partial_function
 } ,
  morekeywords=[2]{rec, return, bind, foreach, if, then, else, and, do, let, in, res, spec, fail, assert, while, case, of},
  sensitive=True,
  morecomment=[s]{(\*}{\*)},
}

\graphicspath{{./figures/}}

\newcommand{\btree}{B$^+$-tree}
\newcommand{\btrees}{B$^+$-trees}

\begin{document}

\title{A Verified Implementation of B$^+$-trees in Isabelle/HOL}
%
%
\author{Niels Mündler\inst{1}\orcidID{0000-0003-3851-2557} \and
Tobias Nipkow\inst{2}\orcidID{0000-0003-0730-515X}}
\authorrunning{N. Mündler \and T. Nipkow}
%
\institute{Department of Computer Science, ETH Zurich, Switzerland
\email{nmuendler@ethz.ch} \and
Department of Informatics, Technical University of Munich, Germany
\url{https://in.tum.de/~nipkow/}}
%
%
%
%

\maketitle

\begin{abstract}
    In this paper we present the verification of an imperative
    implementation of the ubiquitous \btree\ data structure in the
    interactive theorem prover Isabelle/HOL. The implementation supports
    membership test, insertion and range queries with efficient binary
    search for intra-node navigation. The imperative implementation is
    verified in two steps: an abstract set interface is refined to an
    executable but inefficient purely functional implementation which is further
    refined to the efficient imperative implementation.
\end{abstract}

\section{Introduction}
\label{sec:introduction}

\btrees\ form the basis of virtually all modern relational database management systems (RDBMS)
and file systems.
Even single-threaded databases are non-trivial to analyse and verify,
especially machine-checked.
Meanwhile it is important to verify various properties like functional correctness,
termination and runtime,
since RDBMS are ubiquitous and employed in critical contexts,
like the banking sector and realtime systems.
The only work in the literature on that topic that we are aware of
is the work by Malecha \emph{et al.} \cite{DBLP:conf/popl/MalechaMSW10}.
However, it lacks the commonly used range
query operation, which returns a pointer to the lower bound of a given value
in the tree
and allows to iterate over all successive values.
This operation is particulary challenging to verify as it requires
to mix two usually strictly separated abstractions of the tree
in order to reason about its correctness.
We further generalize the implementation of node internal
navigation.
This allows to abstract away from its implementation
and simplifies proofs.
It further allows us to supply an implementation of
efficient binary search, a practical and widespread runtime improvement
as nodes usually have a size of several kilobytes.
We provide a computer assisted proof in the interactive
theorem prover Isabelle/HOL \cite{DBLP:books/sp/NipkowK14} for the functional
correctness of an imperative implementation of the \btree\ data-structure
and present how we dealt with the resulting technical verification challenges.

\section{Contributions}
\label{sec:contributions}

In this work, we specify the \btree\ data structure in the
functional modeling language higher-order logic (HOL).
The tree is proven to refine a finite set of linearly ordered elements.
All proofs are machine-checked in the Isabelle/HOL framework.
Within the framework,
the functional specifications already yield automatic extraction of executable,
but inefficient code.

 The contributions of this work are as follows
\begin{itemize}
   \item The first verification of genuine range queries,
         which require additional insight in refinement over iterating over the whole tree.
   \item The first efficient intra-node navigation based on binary rather than linear search.
\end{itemize}

The remainder of the paper is structured as follows.
In \cref{sec:related_work}, we present a brief overview on related
work.
The definition of \btree\ and our approach is introduced in \cref{sec:approach}.
In \cref{sec:set,sec:range},
we refine a functionally correct, abstract specification of
point, insertion and range queries as well as iterators
down to efficient imperative code.
Finally, we present learned lessons and evaluate the results
in \cref{sec:conclusion}.

The complete source code of the implementation referenced in this research
is accessible via the Archive of Formal Proofs \cite{DBLP:journals/afp/Mundler21}.

\subsection{Related Work}
\label{sec:related_work}

There exist two pen and paper verifications of \btree\ implementations via a rigorous formal approach.
Fielding \cite{Fielding80} uses gradual refinement of abstract
implementations.
Sexton and Thielecke \cite{DBLP:journals/entcs/SextonT08} show how to use 
separation logic in the verification.
These are more of a conceptual guideline on approaching a fully machine checked proof.

There are two machine checked proofs of imperative implementations.
In the work of Ernst \emph{et al.} \cite{DBLP:journals/sosym/ErnstSR15},
an imperative implementation is directly verified
by combining interactive theorem proving in KIV \cite{ReifKIV}
with shape analysis using TVLA \cite{DBLP:journals/toplas/SagivRW02}.
The implementation lacks shared pointers between leaves.
This simplifies the proofs about tree invariants.
However, the tree therefore also lacks iterators over the leaves,
and the authors present no straightforward solution to implement them.
Moreover, by directly verifying an imperative version only,
it is likely that small changes in the implementation will
break larger parts of the proof.

Another direct proof on an imperative implementation
was conducted by Malecha~\emph{et~al.}~\cite{DBLP:conf/popl/MalechaMSW10}, with the Ynot
extension to the interactive theorem prover Coq.
Both works use recursively defined shape predicates
that describe formally how the nodes and pointers
represent an abstract tree of finite height.
The result is both a fairly abstract specification of a \btree,
that leaves some design decisions to the imperative implementation,
and an imperative implementation that supports
iterators.

Due to the success of this approach,
we follow their example and define these predicates functionally.
One example of the benefits of this approach is that we were able
to derive finiteness and acyclicity only from the
relation between imperative and functional specification.
In contrast to previous work, the functional predicates describing the tree shape are kept
completely separated from the imperative implementation,
yielding more freedom for design choices within the imperative refinement.
Both existing works rely on linear search for intra-node navigation,
which we improve upon by providing binary search.
We extend the extraction of an iterator
by implementing an additional range query operation.

\section{\btrees\ and Approach}
\label{sec:approach}

The \btree\ is a ubiquitous data structure to efficiently retrieve and manipulate
indexed data stored on storage devices with slow memory access \cite{DBLP:journals/csur/Comer79}.
They are $k$-ary balanced search trees, where $k$ is a free parameter.
We specify them as implementing a set interface on elements of type \emph{'a},
where the elements in the leaves comprise the content of an abstract set.
The inner nodes contain separators.
These have the same type \emph{'a} as the set content,
but are only used to guide the recursive navigation through the tree
by bounding the elements in the neighboring subtrees.
Further the leaves usually contain pointers
to the next leaf, allowing for efficient iterators and range queries.
A more formal and detailed outline of \btrees\ can be found in \cref{sec:data_structure_defs}.

The goal of this work is to define this data structure
and implement and verify efficient heap-based imperative operations on them.
For this purpose, we introduce a functional, algebraic definition and
specify all invariants on this level that can naturally be expressed in the algebraic domain.
It is important to note that this representation is not complete,
as aliased pointers are left out on the algebraic level.
However, important structural invariants, such as sortedness and balancedness
can be verified.

In a second step an imperative definition is introduced,
that takes care of the refinement of lists to arrays in the heap
and introduces (potentially shared) pointers instead of algebraic structures.
Using a refinement relationship, we can prove that an imperative refinement
of the functional specification preserves the structural invariants
of the imperative tree on the heap.
The only remaining proof obligation on this level is to ensure the correct linking
between leaf pointers.

The above outlined steps are performed via manual refinement in Imperative HOL \cite{DBLP:conf/tphol/BulwahnKHEM08}.
We build on the library of verified imperative utilities
provided by the Separation Logic Framework \cite{DBLP:journals/afp/LammichM12}
and the verification of B-trees \cite{DBLP:journals/afp/Mundler21},
namely list interfaces and partially filled arrays.
The implementation is defined with respect to an abstract imperative
operation for node-internal navigation.
This means that within each node, we do not specify
how the correct subtree for recursive queries is found,
but only constrain some characteristics of the result.
We provide one such operation that employs linear search,
and one that conducts binary search.
All imperative programs are shown to refine the functional specifications
using the separation logic utilities from the Isabelle Refinement Framework by
Lammich \cite{DBLP:journals/jar/Lammich19}.

\subsection{Notation}

Isabelle/HOL conforms to everyday mathematical notation for the most part.
For the benefit of the reader who is unfamiliar with Isabelle/HOL, we establish
notation and in particular some essential datatypes together with their primitive
operations that are specific to Isabelle/HOL. We write \textit{t :: 'a} to specify that
the term \textit{t} has the type \textit{'a} and \textit{'a $\Rightarrow$ 'b}
for the type of a total function from \textit{'a} to \textit{'b}.
The type for natural numbers is \textit{nat}.
Sets with elements of type \textit{'a} have the type \textit{'a set}.
Analogously, we use \textit{'a list} to describe lists, which are constructed as the empty
list \textit{[]} or with the infix constructor \textit{\#}, and are appended with the infix operator
\textit{@}. The function \textit{concat} concatenates a list of lists.
The function \textit{set} converts a list into a set. For optional values, Isabelle/HOL
offers the type \textit{option} where a term \textit{opt :: 'a option} is either \textit{None} or \textit{Some a}
with \textit{a :: 'a}.

\subsection{Definitions}
\label{sec:data_structure_defs}

We first define an algebraic version of \btrees\ as follows:

\begin{lstlisting}[mathescape=true, language=Isabelle,label=lst:btree-def]
datatype 'a bplustree =
    Leaf ('a list) |
    Node (('a bplustree \<times> 'a ) list) ('a bplustree)
\end{lstlisting}

\begin{figure}
    \centering
    \includegraphics[width=0.5\linewidth]{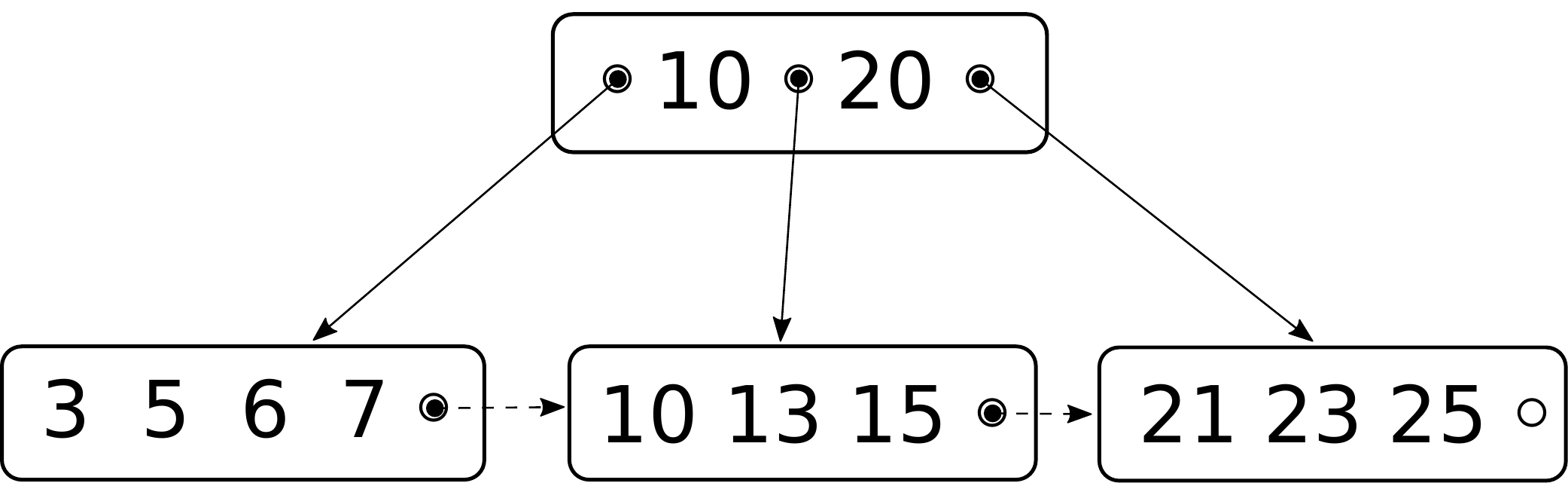}
    \caption[Visualization of a \btree]
    {Nodes contain several elements, the internal list/array structure is not depicted.
    The dotted lines represent links to following leaf nodes that are not present in the algebraic formulation.}
    \label{fig:btree-basic}
\end{figure}

Every node \emph{Node} [($t_1$,$a_1$), …, ($t_n$,$a_n$)] $t_{n+1}$ contains an interleaved list of \textit{keys} or \emph{separators} $a_i$ and \textit{subtrees} $t_i$.
We write as $t_i$ the subtree to the left of $a_i$ and
$t_{i+1}$ the subtree to the right of $a_i$.
We refer to $t_{n+1}$ as the \textit{last} subtree.
The leaves \emph{Leaf} [$v_1$, …, $v_n$] contain a list of \textit{values} $v_i$.
The concatenation of lists of values of a tree $t$ yields
all elements contained in the tree. We refer to this list as \emph{leaves t}.
A \btree\ with the above structure must fulfill the invariants
\textit{balancedness}, \textit{order} and \textit{alignment}.

\textit{Balancedness} requires
that each path from the root to a leaf has the same length.
In other words, the height of all trees in one level of the tree must be equal,
where the height is the maximum path length to a leaf.

The \textit{order} property ensures a minimum and maximum
number of subtrees for each node.
A \btree\ is of order $k$, if each internal node has at least $k+1$
subtrees and at most $2k+1$.
The root is required to have a minimum of 2 and a maximum of $2k+1$ subtrees.
We require that $k$ be strictly positive, as for $k = 0$ the requirements on the tree
root are contradictory.

\textit{Alignment} means that keys are sorted with respect to separators:
For a separator $k$ and all keys $l$ in the subtree to the left, $l < k$,
and all keys $r$ in the subtree to the right, $k \leq r$.
(where $\leq$ and $<$ can be exchanged).

For the values within the leaves, \textit{sortedness} is required explicitly.
We require the even stronger fact that \emph{leaves t} is sorted.
This is a useful statement when arguing about the correctness of set operations.

\subsection{Implementation Definitions}
\label{sec:data_structure_imps}

Proofs about the correctness of operations
with respect to implementing an abtract set interface and
preserving these invariants
are only done on the abstract level, where they are much simpler
and many implementation details can be disregarded.
It will serve as a reference point for the efficient
imperative implementation.

The more efficient executable implementation of \btrees\ is defined
on the imperative level.
Each imperative node contains non-null pointers (\emph{ref}) rather than the algebraic subtree.
We refine lists with partially filled arrays of capacity $2k$.
A partially filled array $(a,n)$ with capacity $c$ is an array $a$ of fixed size $c$.
The array consists of the elements at indices 0 to $n-1$.
Element accesses beyond index $n$ are undefined.
Unlike dynamic arrays, partially filled arrays are not expected to grow or shrink.
Each imperative node contains the equivalent information to an abstract node.
The only addition is that leaves now also contain a pointer to another leaf,
which will form a linked list over all leaves in the tree.
This was not implemented in the algebraic version as it requires pointer aliasing.

\begin{lstlisting}[mathescape=true, language=Isabelle,label=lst:btree-imp-def]
datatype 'a btnode =
  Btleaf ('a pfarray) ('a btnode ref option) |
  Btnode (('a btnode ref option \<times> 'a) pfarray) ('a btnode ref) 
\end{lstlisting}

In order to use the algebraic data structure as a reference point,
we introduce a refinement relation.
The correctness of operations on the imperative node
can then be shown by relating imperative input and output
and to the abstract input and output of a correct abstract operation.
In particular we want to show that if we assume \emph{R t $t_i$},
where $R$ is the refinement relation and $t$ and $t_i$ are the abstract
and the imperative version of the same conceptual tree,
\emph{R o(t) $o_i$($t_i$)} should hold, where $o_i$ is the imperative refinement
of operation $o$.
The relation is expressed as a separation logic formula that links an abstract tree to its
imperative equivalent.

The notation for separation logic in Isabelle is quickly summarized in the list below.
\begin{itemize}
    \item \textit{emp} holds for the empty heap
    \item \textit{true} and \textit{false} hold for every and no heap respectively
    \item $\uparrow(P)$ holds if the heap is empty and predicate $P$ holds
    \item $a \mapsto_r x$ holds if the heap at location $a$ is reserved and contains
    value $x$
    \item $\exists_A\ x.\ P\ x$ holds if there exists some $x$ such that $P x$
    holds on the heap.
    \item $P_1 * P_2$ denotes the separating conjunction and holds if each assertion $P_1$ and $P_2$ hold on non-overlapping parts
    of the heap
    \item \emph{is\_pfa c xs xsi} holds if $xsi$ is a partially filled array
    with capacity $c$ and \emph{xs[i] = xsi[i]} holds for all $i \leq |xs| = |xsi|$.
    \item \emph{list\_assn R xs ys} holds if \emph{R xs[i] ys[i]} holds for all $i \leq |xs| = |ys|$.
\end{itemize}
Separation Logic formulae express assertions that can be made about the state of the heap.
They are therefore just called \emph{assertion} in the following.
The assertion $P$ describes all heaps for which the formula $P$ evaluates to \emph{true}.
The entailment $P \Longrightarrow_A Q$ holds iff $Q$ holds in every heap in which $P$ holds.
For two assertions $P$ and $Q$, $P = Q$ holds iff $P \Longrightarrow_A Q \wedge Q \Longrightarrow_A P$.
For proving imperative code correct, assertions are used in the context of Hoare triples.
We write $\langle P\rangle\ c\ \langle\lambda r.\ Q\ r\rangle$ if, for any heap where $P$ holds, after executing
imperative code $c$ that returns value $r$, formula $Q\ r$ holds on the resulting heap.
$\langle P\rangle\ c\ \langle\lambda r.\ Q\ r\rangle_t$ is a shorthand for $\langle P\rangle\ c\ \langle\lambda r.\ Q\ r * \mathit{true}\rangle$
More details can be found in the work of Lammich and Meis \cite{DBLP:journals/afp/LammichM12}.

The assertion \emph{bplustree\_assn} expresses the refinement relation.
It relates an algebraic tree (\emph{bplustree})
and a non-null pointer to an imperative tree $a$ (\emph{btnode ref}), pinning its first leaf $r$ and the first leaf of the next sibling $z$.
The formal relation is shown in \cref{fig:btree-assn}.

\begin{figure}
   \centering 
\begin{lstlisting}[mathescape=true, language=Isabelle,label=lst:btree-relation]
fun bplustree_assn :: nat $\Rightarrow$ 'a bplustree $\Rightarrow$ 'a btnode ref
      $\Rightarrow$ 'a btnode ref option $\Rightarrow$ 'a btnode ref option where
    bplustree_assn k (Node ts t) a r z = $\exists_A$ tsi ti tsi' rs.
      a $\mapsto_r$ Btnode tsi ti
      -- $\text{\emph{Obtain list with array contents for folding list\_assn}}$
    * is_pfa (2*k) tsi' tsi
    * $\uparrow$(length tsi' = length rs)
      -- $\text{\emph{Recursively apply the assertion to subtree pointers}}$
    * list_assn (($\lambda$ t (ti,r',z'). bplustree_assn k t (the ti) r' z') $\times_a$ id_assn) ts (
      -- $\text{\emph{Pointers to left/right sibling are obtained by offset zipping}}$
        zip (zip (map fst tsi') (zip (butlast (r#rs)) rs))) (map snd tsi')))
    * bplustree_assn k t ti (last (r#rs)) z)
  | bplustree_assn k (Leaf xs) a r z = $\exists_A$ xsi fwd. 
        a $\mapsto_r$ Btleaf xsi fwd * is_pfa (2*k) xs xsi * $\uparrow$(fwd = z) * $\uparrow$(r = Some a)
\end{lstlisting}
    \caption[Assertion describing the imperative \btree]{
        The \btree\ is specified by the split factor $k$, an abstract tree,
        a pointer to its root $a$, a pointer to its first leaf $r$ and a pointer
        to the first leaf of the next sibling $z$.
        The pointers to first leaf and next first leaf are used
        to establish the linked leaves invariant.
    }
    \label{fig:btree-assn}
\end{figure}

The main structural relationship between abstract and imperative tree
is established by linking abstract list and array via the \textit{is\_pfa} predicate.
We then fold  over the two lists using \textit{list\_assn},
which establishes a refinement relation for every pair of list elements.

In addition to the refinement of data structures,
the first leaf $r$ and next leaf $z$ are used to express the structural invariant
that the leaves are correctly linked.
There is no abstract equivalent for the forwarding pointers in the leaves,
therefore we only introduce and reason about their state on the imperative layer.
The invariant is ensured by passing the first leaf of the right neighbor to each subtree.
The pointer is passed recursively to the leaf node,
where it is compared to the actual pointer of the leaf.
All of this happens in the convoluted \textit{list\_assn}, by
folding over the list of the leaf pointer list \textit{rs} zipped with itself, offset by one.
The linking property is required for the iterator on the tree in \cref{sec:imperative_iter}.

\subsection{Node internal navigation}
\label{sec:split}

In order to define meaningful operations that navigate
the node structure of the \btree,
we need to find a method that handles search within a node.
Ernst \emph{et al.} \cite{DBLP:journals/sosym/ErnstSR15} and Malecha~\emph{et~al.}~\cite{DBLP:conf/popl/MalechaMSW10}
both use a linear search through the key and value lists.
However, \btrees\ are supposed to have memory page sized nodes \cite{DBLP:journals/csur/Comer79}, 
which makes a linear search impractical.

We introduce a context (\emph{locale} in Isabelle) in which we assume that we
have access to a function that correctly navigates through the node internal structure.
\emph{Correct} in this context meaning that the selected subtree for recursive calls
will lead to the element we are looking for.
We call this function \emph{split}, and define it only by its behavior.
The specification for \emph{split} is given in \cref{fig:split-def} (where \emph{'b = 'a bplustree $\times$ 'a}).
A corresponding function \emph{split\_list} is defined on the separator-only lists in the leaf nodes.

\begin{figure}
   \centering 
\begin{lstlisting}[mathescape=true, language=Isabelle,label=lst:split-def]
locale split_tree =
    fixes split ::  "'b list \<Rightarrow> 'a \<Rightarrow> 'b list \<times> 'b list
    "split xs p = (ls,rs) \<Longrightarrow> xs = ls @ rs" 
    "split xs p = (ls@[(sub,sep)],rs); sorted_less (separators xs) \<Longrightarrow> sep < p" 
    "split xs p = (ls,(sub,sep)#rs); sorted_less (separators xs) \<Longrightarrow> p \<le> sep" 

\end{lstlisting}
    \caption[Definition of \emph{split}]{
        Given a list of separator-subtree pairs and a search value $x$, the function should return the pair $(s,t)$ such that,
        according to the structural invariant of the \btree, $t$ must contain $x$ or will hold $x$ after a correct insertion.
    }
    \label{fig:split-def}
\end{figure}

In the following sections, all operations are defined and verified
based on \emph{split} and \emph{split\_list}.
When approaching imperative code extraction,
we provide a binary search based imperative function, that refines \emph{split}.
Thus we obtain imperative code that makes use of an efficient
binary search, without adding complexity to the proofs.
The definition and implementation closely follows
the approach described in detail in the
verification of B-trees \cite{DBLP:journals/afp/Mundler21}.

\section{Set operations}
\label{sec:set}

\btrees\ refine sets on linearly ordered elements.
For a tree $t$, the refined abstract set is computed as \emph{set (leaves t)}.
The set interface requires that there should be query, insertion and deletion
operations $o_t$ such that \emph{set (leaves ($o_t$ t)) = o (set (leaves t))}.
Moreover, the invariants described in \cref{sec:approach}
can be assumed to hold for $t$ and are required for $o_t$.
We provide these operations and show their correctness on the functional
layer first, then refine the operations further to the imperative
layer.
For point queries and insertion, we follow the implementation
suggested by Bayer and McCreight \cite{DBLP:journals/acta/BayerM72}.

\subsection{Functional Point Query}
\label{sec:functional_pq}

For an inner node $t$ and a searched value $x$, find the correct subtree $s_t$
such that if a leaf of $t$ contains $x$, a leaf of $s_t$ must contain $x$.
Then recurse on $s_t$.
Inside the leaf node, we search directly in the list of values.
We make use of the \textit{split} and \textit{isin\_list} operation,
as described in \cref{sec:split}.

\begin{lstlisting}[mathescape=true, language=Isabelle,label=lst:isin-def]
fun isin:: "'a bplustree $\Rightarrow$ 'a $\Rightarrow$ bool" where
  isin (Leaf ks) x = (isin_list x ks) |
  isin (Node ts t) x = (case split ts x of
     (_,(sub,sep)#rs) $\Rightarrow$ isin sub x
   | (_,[]) $\Rightarrow$ isin t x
  )
\end{lstlisting}

Since this function does not modify the tree involved at all,
we only need to show that it returns the correct value.

\begin{lstlisting}[mathescape=true, language=Isabelle,label=lst:isin-set-inorder]
theorem assumes "sorted_less (leaves t)" and "aligned l t u" 
  shows "isin t x = (x $\in$ set (leaves t))"
\end{lstlisting}

In general, these proofs on the abstract level are 
based on yet another refinement relation suggested by Nipkow~\cite{DBLP:conf/itp/Nipkow16}. 
We say that the \btree\ $t$ refines a sorted list of its leaf values, \emph{leaves t},
the concatenated lists of values in leafs visited in in-order traversal of the tree.
We argue that recursing into a specific subtree
is equivalent to splitting this list at the correct position
and searching in the correct sublist.
The same approach was applicable for proving the correctnes of functional
operations on B-trees \cite{DBLP:journals/afp/Mundler21}.

The proofs on the functional level can therefore be made concise.
We go on and define an imperative version of the operation that
refines each step of the abstract operation to equivalent operations on the imperative tree.

\subsection{Imperative Point Query}
\label{sec:imperative_pq}

The imperative version of the point query is a partial function.
Termination cannot be guaranteed anymore,
at least without further assumptions.
This is inevitable since the function would not terminate
given cyclic trees.
However, we will show that if the input refines an abstract tree,
the function terminates and is correct.
The imperative \emph{isin$_i$} refines each step of the abstract
operation with an imperative equivalent.
The result can be seen in \cref{fig:isin-imp-def}.

\begin{figure}
    \centering
\begin{lstlisting}[mathescape=true, language=Isabelle,label=lst:isin-imp-def]
partial_function (heap) isin$_i$ :: "'a btnode ref $\Rightarrow$ 'a $\Rightarrow$  bool Heap" where
  isin$_i$ p x = do {
  node $\leftarrow$ !p;
  (case node of
     Btleaf xs _ $\Rightarrow$ isin_list$_i$ x xs |
     Btnode ts t $\Rightarrow$ do {
       i $\leftarrow$ split$_i$ ts x;
       tsl $\leftarrow$ length ts;
       if i < tsl then do {
         s $\leftarrow$ get ts i;
         let (sub,sep) = s in
           isin$_i$ (the sub) x
       } else
           isin$_i$ t x
    }
)}
\end{lstlisting}
\caption[Definition of \emph{isin$_i$}]{
    The imperative refinement of the \emph{isin} function.
    As a partial function, its termination is not guaranteed for all inputs.
    Additionally it implicitly makes use of the heap monad.
}
\label{fig:isin-imp-def}
\end{figure}

Again, we assume that \emph{split$_i$} performs the correct node internal search
and refines an abstract \emph{split}.
Note how \emph{split$_i$} does not actually split
the internal array, but rather returns the index of the pair
that would have been returned by the abstract split function.
The pattern matching against an empty list
is replaced by comparing the index to the length of the list $l$.
In case the last subtree should be recursed into, the whole list $l$ is returned.

In order to show that the function returns the correct result,
we show that it performs the same operation on the imperative tree
as on the algebraic tree.
This is expressed in Hoare triple notation and separation logic.

\begin{lstlisting}[mathescape=true, language=Isabelle,label=lst:isin-refines]
lemma assumes "k > 0" and "root_order k t" and "sorted_less (inorder t)"
   and "sorted_less (leaves t)" shows
   "\<langle>bplustree_assn k t ti r z\<rangle>
     isin$_i$ ti x
   \<langle>$\lambda$y. bplustree_assn k t ti r z * $\uparrow$(isin t x = y)\<rangle>$_t$"
\end{lstlisting}

The proof follows inductively on the structure of the abstract tree.
Assuming structural soundness of the abstract tree refined by the pointer passed in,
the returned value is equivalent to the return value of the abstract function.
We must explicitly show that the tree on the heap
still refines the same abstract tree after the operation,
which was implicit on the abstract layer.
It follows directly, since no operation in the imperative
function modifies part of the tree.

\subsection{Insertion and Deletion}
\label{sec:insert_delete}

The insertion operation and its proof of correctness largely line up with the one for point queries.
But since insertion modifies the tree,
we need to additionally show on the abstract level that the modified tree
maintains the invariants of \btrees.

On the imperative layer, we show that the heap state
after the operation refines the tree
after the abstract insertion operation.
It follows that the imperative operation
also maintains the abstract invariants.
Moreover, we need to show that the linked list 
among the leaf pointers is correctly maintained throughout the operation.
This can only be shown on the imperative level as there is no abstract equivalent
to the shared pointers.

\begin{lstlisting}[mathescape=true, language=Isabelle,label=lst:insert-refines]
lemma assumes "k > 0" and "sorted_less (inorder t)"
    and "sorted_less (leaves t)" and "root_order k t" shows
    "\<langle>bplustree_assn k t ti r z\<rangle>
    insert$_i$ k x ti
    \<langle>\<lambda>u. bplustree_assn k (insert k x t) u r z\<rangle>$_t$"
\end{lstlisting}

We provide a verified functional definition of deletion and a definition of an imperative refinement.
Showing the correctness of the imperative version would largely follow
the same pattern as the proof of the correctness of insertion.
The focus of this work is not on basic tree operations, but on obtaining a (range) iterator view on the tree.

\section{Range operations}
\label{sec:range}

This section introduces both how the general iterator
on the tree leaves is obtained and the technical challenges involved
(\cref{sec:imperative_iter})
as well as how to obtain an iterator on a specific
subset of elements efficiently (\cref{sec:imperative_range}).

On the functional level, the forwarding leaf pointers in each leaf
are not present, as this would require aliasing.
Therefore, the abstract equivalent of an iterator
is a concatenation of all leaf contents.
When refining the operations, we will make use of the leaf pointers
to obtain an efficient implementation.

\subsection{Iterators}
\label{sec:imperative_iter}

To obtain an iterator, recurse down the tree to obtain the first leaf.
From there we follow leaf pointers until we reach the final leaf marked by a null forwarding pointer.
From an assertion perspective the situation is more complex.
Recall the refinement relation between abstract and implemented \btree.
It is important to find an explicit formulation of the linked list view on the leaf pointers.
Meanwhile, we want to ensure that the complete tree does not change by iterating through the leaves.
We cannot express an assertion about the linked list along the leaves
and the assertion on the whole tree in two fully independent predicates
as the memory described overlaps.
Separation logic forces us to not make statements about the contents of
any memory location twice.

\begin{figure}
    \centering
    \includegraphics[width=1\linewidth]{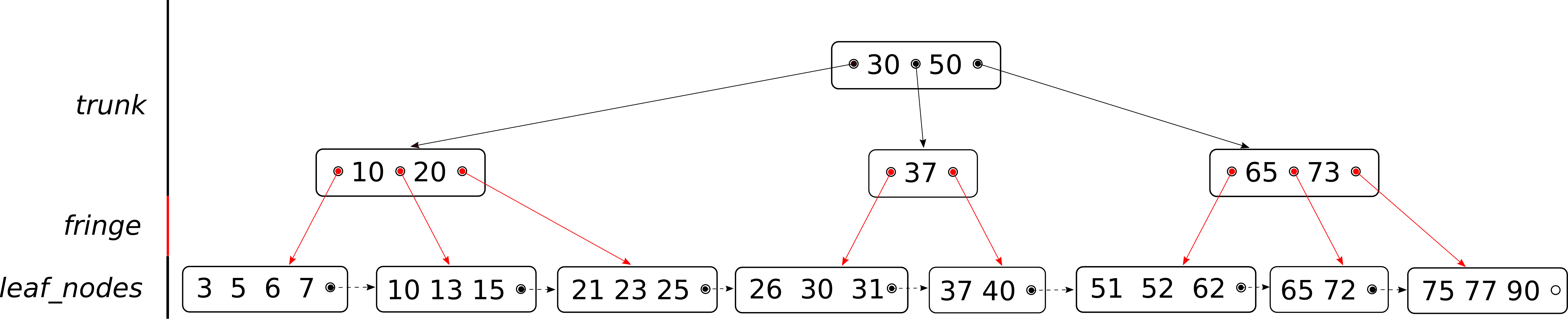}
    \caption[Split view of the \btree]
    {In order to obtain separate assertions about the concatenated leaf list (\emph{leaf\_nodes})
    and the internal nodes (\emph{trunk}) of the tree, the structure is abstractly split along the
    pointers marked in red, the \emph{fringe}. In order to be able to combine the \emph{leaf\_nodes} and the \emph{trunk} together,
    the \emph{fringe} has to be extracted and shared explicitly.}
    \label{fig:btree-view-split}
\end{figure}

We follow the approach of Malecha \emph{et al.} \cite{DBLP:conf/popl/MalechaMSW10} and
try to find an equivalent formulation that separates the whole tree in a
view on its inner nodes and the linked leaf node list.
The central idea to separate the tree is to
express that the linked leaf nodes refine \emph{leaf\_nodes t}
and that the inner nodes refine \emph{trunk t}, as depicted in \cref{fig:btree-view-split}.
These are two independent parts of the heap and therefore
the statements can be separated using the separating conjunction.

Formally, we define an assertion \emph{trunk\_assn} and \emph{leaf\_nodes\_assn}.
The former is the same as \emph{bplustree\_assn} (see \cref{fig:btree-assn}),
except that we remove all assertions about the content of the tree in the \emph{Leaf} case.
The latter is defined similar to a linked list refining a list of abstract tree leaf nodes,
shown in \cref{fig:leaf-nodes-assn}.
The list is refined by a pointer to the head of the list,
which refines the head of the abstract list.
Moreover, the imperative leaf contains a pointer to the next element in the list.

With these definitions, we can show that the heap describing the imperative tree may be
split up into its leaves and the trunk.

\begin{lstlisting}[mathescape=true, language=Isabelle,label=lst:btree-view-split-oneway]
lemma "bplustree_assn k t ti r z
    $\Longrightarrow_A$ leaf_nodes_assn k (leaf_nodes t) r z * trunk_assn k t ti r z"
\end{lstlisting}

\begin{figure}
    \centering
\begin{lstlisting}[mathescape=true, language=Isabelle,label=lst:leaf-nodes-assn]
fun leaf_nodes_assn where
  "leaf_nodes_assn k ((Leaf xs)#lns) (Some r) z =
 (\<exists>$_A$ xsi fwd.
      r \<mapsto>$_r$ Btleaf xsi fwd
    * is_pfa (2*k) xs xsi
    * leaf_nodes_assn k lns fwd z
  )" |
  "leaf_nodes_assn k [] r z = \<up>(r = z)" |
  "leaf_nodes_assn _ _ _ _ = false"
\end{lstlisting}
\caption[Definition of \emph{leaf\_nodes\_assn}]{
    The refinement relation for leaf nodes comprises the refinement
    of the node content as well as the recursive property of linking correctly to the next node.
}
\label{fig:leaf-nodes-assn}
\end{figure}

However, we cannot show that a structurally consistent, unchanged \btree\
is still described by the combination of the two predicates.
The reason is that we cannot express that the linked leaf nodes
are precisely the leaf nodes on the lowest level of the trunk, depicted
in red in \cref{fig:btree-view-split}.

The root of this problem is actually a feature of the refinement approach.
When stating that a part of the heap
refines some abstract data structure,
we make no or little statements about concrete memory locations or pointers.
This is useful, as it reduces the size of the specification
and the proof obligations.
In this case we need to find a way around it.


We need to specifically express that the leaf pointers,
and not the abstract structure they refine,
are precisely the same in the two statements.

In a second attempt, the sharing is made explicit.
We extract from the whole tree the precise list of pointers to leaf nodes, the \emph{fringe}
in the correct order.
Recursively, the fringe of a tree is the concatenation of all fringes
in its subtrees.
The resulting assertion, taking the fringe into account, can be seen in \cref{fig:btree-assn-leaves}.
As a convenient fact, this assertion is equivalent to \cref{fig:btree-assn}.

\begin{lstlisting}[mathescape=true, language=Isabelle,label=lst:btree-extract-fringe]
lemma bplustree_extract_fringe:
    "bplustree_assn k t ti r z = ($\exists_A$fringe. bplustree_assn_fringe k t ti r z fringe)"
\end{lstlisting}

\begin{figure}
    \centering
\begin{lstlisting}[mathescape=true, language=Isabelle,label=lst:btree-assn-leaves]
fun bplustree_assn_fringe where
    bplustree_assn_fringe k (Leaf xs) a r z fringe =
    $\exists_A$ xsi fwd.
        a $\mapsto_r$ Btleaf xsi fwd
      * is_pfa (2*k) xs xsi
      * $\uparrow$(fwd = z)
      * $\uparrow$(r = Some a)
      -- $\text{\emph{In case of a singleton leaf, the leaf itself is the fringe of the tree}}$
      * $\uparrow$(fringe = [a])
    |
    bplustree_assn_fringe k (Node ts t) a r z fringe =
    $\exists_A$ tsi ti tsi' tsi'' rs fr_sep.
        a $\mapsto_r$ Btnode tsi ti
      * is_pfa (2*k) tsi' tsi
      * $\uparrow$(length tsi' = length rs)
      -- $\text{\emph{The fringe is decomposed into the fringe of each subtree}}$
      * $\uparrow$(concat fr_sep = fringe)
      * $\uparrow$(length fr_sep = length rs + 1)
      -- $\text{\emph{Folding over all subtrees as before, now passing each subfringe to subtrees}}$
      * list_assn (
          ($\lambda$ t (ti,r',z',fr). bplustree_assn_fringe k t (the ti) r' z' fr)
           $\times_a$ id_assn
        ) ts (zip 
            (zip (map fst tsi') (zip (butlast (r#rs)) (zip rs (butlast fr_sep))))
            (map snd tsi')
        )
      * bplustree_assn_fringe k t ti (last (r#rs)) (last (rs@[z])) (last fr_sep)
       
\end{lstlisting}
    \caption[\btree\ assertion with extracted fringe]{
        An extended version of the \btree\ assertion from \cref{fig:btree-assn} on imperative tree
        root $a$, first leaf $r$, first leaf of the next sibling $z$ and leaf pointer list $fringe$.
        In order to be able to correctly relate leaf view and internal nodes,
        the shared pointers \emph{fringe} are made explicit, without accessing their memory location.
    }
    \label{fig:btree-assn-leaves}
\end{figure}

Using the \emph{fringe}, we can precisely state an equivalent separated assertion.
We describe the trunk with the assertion \emph{trunk\_assn},
which is the same as \emph{bplustree\_assn\_fringe},
except that the \emph{Leaf} case is changed to only $\uparrow(r = \mathit{Some}\ a \wedge \mathit{fringe} = [a])$.
In addition, we extend the definition of \emph{leaf\_nodes\_assn}
to take the \emph{fringe} pointers into account.
We now require that the \emph{fringe} of the trunk is
precisely the list of pointers in the linked list refining \emph{leaf\_nodes}.

\begin{lstlisting}[mathescape=true, language=Isabelle,label=lst:btree-view-split]
lemma bplustree_view_split:
  "bplustree_assn_fringe k t ti r z fringe =
   leaf_nodes_assn k (leaf_nodes t) r z fringe * trunk_assn k t ti r z fringe"
\end{lstlisting}

To obtain an iterator on the leaf nodes of the tree,
we obtain the first leaf of the tree.
By the formulation of the tree assertion, we can express
the obtained result using the assertion about the complete tree.

\begin{lstlisting}[mathescape=true, language=Isabelle,label=lst:btree-first-leaf]
lemma assumes "k > 0" and "root_order k t" shows
  "\<langle>bplustree_assn k t ti r z\<rangle>
  first_leaf ti
  \<langle>$\lambda$u. bplustree_assn k t ti r z * $\uparrow$(u = r)\<rangle>$_t$"
\end{lstlisting}

On the result, we can apply lemmas \hyperref[lst:btree-extract-fringe]{\emph{bplustree\_extract\_fringe}} 
and \hyperref[lst:btree-view-split]{\emph{bplustree\_view\_split}}.
The transformed expression states that
the result of \emph{first\_leaf ti} is a pointer to \emph{leaf\_nodes t}.
The tree root \emph{t} remains to refine \emph{trunk t}.

From here, we could define an iterator over the leaf nodes
along the fringe, refining the abstract list \emph{leaf\_nodes}.
Our final goal is to iterate over the values within each array inside the nodes.
We introduce a flattening iterator for this purpose.
It takes an outer iterator over a data structure \textit{a} that returns elements of type \textit{b},
and inner iterator over the data structure \textit{b} that returns elements of type \textit{c}.
It returns an iterator over data structure \textit{a}
that returns the concatenated list of elements of type \textit{c}.
The exact implementation of this iterator is left out as a technical detail.

The list iterator interface used is as defined by Lammich \cite{DBLP:conf/itp/Lammich19}
and specifies the following function.
\begin{itemize}
    \item An \emph{init} function that returns the pointer to the head of the list.
    \item A \emph{has\_next} function that checks whether the current pointer is the null pointer.
    \item A \emph{next} function that returns the the array in the current node and its forwarding pointer.
    \item Proofs that we can transform the \emph{leaves\_assn} statement into 
          a leaf iterator statement and vice versa.
\end{itemize}
We implement such an iterator for the linked list of leaf nodes \emph{leaf\_nodes\_iter}
and combine it with the iterator over partially filled arrays
using the flattening iterator to obtain the \emph{leaves\_iter}.

Finally, we want be able to express that the whole tree does not change throughout the iteration.
For this, we need to keep track of both the leaf nodes assertion and the trunk assertion on \emph{t}.
The assertion describing the iterator therefore contains both.
Most parameters to the iterator assertion are static, and express the context of the iterator,
i.e. the full extent of the leaf nodes.
The iterator state $it$ itself is a pair of an iterator state for a partial array,
the current position in that array and its size,
and a pointer to the next leaf and the final leaf.

\begin{lstlisting}[mathescape=true, language=Isabelle,label=lst:btree-tree-iter]
definition "bplustree_iter k t ti r vs it = \<exists>$_A$ fringe.
  leaves_iter fringe k (leaf_nodes t) (leaves t) r vs it *
  trunk_assn k t ti r None fringe"
\end{lstlisting}

Note how all notion of the explicitly shared fringe
has disappeared from the client perspective
as its existence is hidden within the definition of the tree iterator.
We initialize the iterator using the \emph{first\_leaf} operation
and obtain the singleton tree elements with the flattening iterator.

\subsection{Range queries}
\label{sec:imperative_range}

A common use case of \btrees\ is
to obtain all values within a range \cite{DBLP:journals/ftdb/Graefe11}.
We focus on the range bounded from below, \emph{lrange t x = $\{y \in set(t) | y \geq x\}$}.
From an implementation perspective, the operation is similar to the point query operation.
On the leaf level, it returns a pointer to the reached leaf.
This pointer is then interpreted as iterator over the remaining list of linked leaves.
The range bounded from below comprises all values returned by the iterator.
Due to the lack of a linked leaf list in the abstract tree,
the abstract definition explicitly concatenates all values in the subtrees 
to the right of the reached node.

\begin{lstlisting}[mathescape=true, language=Isabelle,label=lst:btree-lrange]
fun lrange:: "'a bplustree $\Rightarrow$ 'a $\Rightarrow$ 'a list" where
    lrange (Leaf ks) x = (lrange_list x ks) |
    lrange (Node ts t) x = (
        case split ts x of (_,(sub,sep)#rs) $\Rightarrow$ (
               lrange sub x @ (concat (map leaves rs)) @ leaves t
        )
     | (_,[]) $\Rightarrow$ lrange t x
    )
\end{lstlisting}
  
As before, we assume that there exists a function \textit{lrange\_list} that
obtains the \emph{lrange} from a list of sorted values.

The verification of the imperative version turns out to be not as straightforward
as expected, exactly due to this recursive step.
The reason is that iterators can only be expressed on a complete tree,
where the last leaf is explicitly a null pointer.
The linked list of a subtree is however bounded by valid leaves,
precisely the first leaf of the next subtree.

In order implement and verify a refinement of this function we therefore decide
to implement an intermediate abstract function \emph{leaf\_nodes\_lrange}.
This function returns the leaf nodes comprising the \emph{lrange} instead of their contents.

\begin{minipage}{\linewidth}
\begin{lstlisting}[mathescape=true, language=Isabelle,label=lst:btree-leaves-range]
fun leaf_nodes_lrange:: 'a bplustree $\Rightarrow$ 'a $\Rightarrow$ 'a bplustree list where
  leaf_nodes_lrange (Leaf ks) x = [Leaf ks] |
  leaf_nodes_lrange (Node ts t) x = case split ts x of
      (_,(sub,sep)#rs) $\Rightarrow$ 
        leaf_nodes_lrange sub x @ concat (map leaf_nodes rs) @ leaf_nodes t
    | (_,[]) $\Rightarrow$ leaf_nodes_lrange t x
  

fun concat_leaf_nodes_lrange where
  concat_leaf_nodes_lrange t x = case leaf_nodes_lrange t x of
  (LNode ks)#list $\Rightarrow$ lrange_list x ks @ concat (map leaves list)
\end{lstlisting}
\end{minipage}

We then show that the concatenation of the contents of the leaf nodes
\emph{concat\_leaf\_nodes\_lrange t x = lrange t x}.
On the imperative layer \emph{leaf\_nodes\_lrange$_i$}
can be obtained using only the \emph{leaf\_nodes} and \emph{trunk}
assertions as we never access the contents of the leaf nodes.
We therefore avoid having to unfold any assertions about the structure of the leaf nodes.
The function returns a pointer that splits the list of leaf nodes of the whole tree,
terminated by the null pointer that marks the end of the complete tree.
We transform the result into an iterator over the leaf nodes,
as this pointer split notation aligns with the definition of \emph{leaf\_nodes\_iter}.
Finally we can transform this and the result of \emph{lrange\_list$_i$} to
an iterator on the singleton leaf elements.

\begin{lstlisting}[mathescape=true, language=Isabelle,label=lst:btree-leaves-range]
lemma assumes "k > 0" and "root_order k t" 
    and "sorted_less (leaves t)" and "Laligned t u" shows 
  "\<langle>bplustree_assn k t ti r None\<rangle>
  concat_leaf_nodes_lrange$_i$ ti x
  \<langle>bplustree_iter k t ti r (lrange t x)\<rangle>$_t$"
\end{lstlisting}


\section{Conclusion}
\label{sec:conclusion}

We were able to formally verify an imperative implementation
of the ubiquitous \btree\ data structure, featuring range queries and binary search.

\subsection{Evaluation}

The \btree\ implemented by Ernst \emph{et al.} \cite{DBLP:journals/sosym/ErnstSR15} features point queries and insertion,
however explicitly leaves out pointers within the leaves,
which forbids the implementation of iterators.
Our work is closer in nature to the \btree\ implementation by Malecha \emph{et al.} \cite{DBLP:conf/popl/MalechaMSW10}.
In addition to the functionality dealt with in their work, we extend
the implementation with a missing Range iterator
and supply a binary search within nodes.
Our approach is modular, allowing for the substitution of parts
of the implementation with even more specialized and sophisticated implementations.

Regarding the leaf iterator, we noticed that in the work of Malecha \emph{et al.}
there is no need to extract the fringe explicitly.
The abstract leaves are defined such that they store the precise heap location of the refining node.
In our proposed definition, the precise heap location is irrelevant in almost every situation and can be omitted.
Only when splitting the tree we obtain the memory location
of nodes explicitly, because these locations are needed to guarantee
structual soundness of the whole tree.
It is hard to quantify or evaluate which approach is more practical.
From a theoretical view point we suggest that a less strict approach restricts the implementation space less
and leaves more design decisions to the specification implementing developer.

With respect to the effort in lines of code and proof
as depicted in \cref{fig:proof-comparison},
our approach is similar in effort to the approach by Malecha \emph{et al.}.
The numbers do not include the newly defined pure ML proof tactics.
It includes the statistics for the additional binary search and range iterator,
that make up around one thousand lines of proof each.

The comparison with Ernst \emph{et al.} is difficult.
Their research completely avoids the usage of linked leaf pointers,
therefore also omitting iterators completely.
The iterator verification makes up a signifant amount of the proof
with at least one thousand lines of proof on its own.
The leaf pointers also affect the verification of point and insertion queries
due to the additional invariant on the imperative level.
We conclude that the Isabelle/HOL framework
provides a feature set
such that verification of \btrees\ is both feasible
and comparable in effort to using Ynot or KIV/TVLA.
The strict separation of a functional and imperative
implementation yields the challenge
of making memory locations explicit where needed.
On the other hand, it permits great freedom
regarding the actual refinement on the imperative level.
\renewcommand*{\thefootnote}{\fnsymbol{footnote}}

\begin{figure}
    \centering
    \begin{tabular}{l|c|c|c}
        \                & Malecha \emph{et al.}\cite{DBLP:conf/popl/MalechaMSW10}$^{+}$ & \cite{DBLP:journals/sosym/ErnstSR15}$^{d}$ & Our approach$^{+}$ \\
        \hline
        Functional code &   360      & -                    & 413  \\ 
        Imperative code &   510      & 1862                  & 1093  \\
        Proofs          &  5190      & 350 + 510 + 2940\footnotemark[1] & 8663 \\
        Timeframe (months) &  ?     & $>6$                      & 6 + 6\footnotemark[2] \\
    \end{tabular}
    \caption[Comparison of (unoptimized) Lines of Code and Proof and time investment in related mechanized \btree\ verifications.]
    {Comparison of (unoptimized) Lines of Code and Proof and time investment in related mechanized \btree\ verifications.
    All approaches are comparable in effort, taking into account implementation specifics.
    The marker $^d$ denotes that the implementation verifies deletion operations, whereas $^+$ denotes the implementation of iterators.
    }
    \label{fig:proof-comparison}
\end{figure}
\footnotetext[1]{
    The proof integrates TVLA and KIV, and hence comprises
    explicitly added rules for TVLA (the first number),
    user-invented theorems in KIV (the second number)
    and "interactions" with KIV (the second number).
    Interactions are i.e. choices of an induction variable, quantifier instantiation
    or application of correct lemmas.
    We hence interpret them as each one apply-Style command and hence
    one line of proof.
}
\footnotetext[2]{
    6 months include the preceding work on the verification
    of B-trees.
    As they share much of the functionality with \btrees\ 
    but required their own specifics,
    the time spent on them cannot be accounted for 1:1.
}

\renewcommand*{\thefootnote}{\arabic{footnote}}
\subsection{Outlook}

This research may serve as a template for the implementation of \btrees\
in Isabelle-LLVM. \cite{DBLP:conf/itp/Lammich19}
At the beginning of this work, the code generator did
not yet support recursive data structures, but
this functionality was added recently.

As of now, the imperative implementation provided by this research was
directly exported into executable imperative code in Haskell, SML and OCaml.
It may thus find applications in the development of
libraries where a verified implementation of a set interface is needed.







\bibliography{ictac2022-bplustrees}

\end{document}